\documentclass[reqno,11pt]{amsart}
\usepackage{epsfig}\usepackage{amsfonts,amsmath,amssymb,mathrsfs,verbatim,amsthm,amscd}
\begin{document}

\title[Superconformal indices, Seiberg dualities and special functions]
{Superconformal indices, Seiberg dualities\\ and special functions%
}

\author{Vyacheslav \,P. Spiridonov}%

\makeatletter
\renewcommand{\@makefnmark}{}
\makeatother
\footnote{\small Talk given at the conference {\em Problems of Theoretical and Mathematical Physics}
dedicated to 110th anniversary of the birth of N.~N.~ Bogolyubov (Moscow and Dubna, 09-13.09.2019),
http://thproxy.jinr.ru/video/bog2019/mp4/13$\_$Spiridonov.mp4\\
\indent
This work is partially supported by the Laboratory of Mirror Symmetry NRU HSE, RF government grant,
ag. no. 14.641.31.0001.}

\address{Bogoliubov Laboratory of Theoretical Physics,
JINR, Dubna, Moscow region, 141980 Russia and
National Research University Higher School of Economics, Moscow, Russia
}

\maketitle

\vspace*{-2em}

\begin{abstract}
This is a brief account of relations between the theory of special functions,
on the one side, and superconformal indices and Seiberg dualities of
four-dimensional $\mathcal{N}=1$ supersymmetric gauge field theories, on the other side.
\end{abstract}

\begin{flushright}
\em To the memory of Richard Askey
\end{flushright}

\vspace*{1em}

{\bf Plain hypergeometric functions.}
Since this is a memorial meeting, I have decided to give a partially historical presentation.
Evidently, history of the subject indicated in the talk title starts from the theory of special functions.
Namely, it is necessary to go back as far as to the times of Isaak Newton, when physics and mathematics
formed one science and they were not separated, as we see it nowadays. Among his numerous great achievements,
in 1665 Newton proved the binomial theorem
\begin{equation}
{}_1F_0(a;x):=\sum_{n=0}^\infty \frac{(a)_n}{n!} x^n=(1-x)^{-a}, \qquad a, x \in\mathbb{C},\quad |x|<1,
\label{binomial}\end{equation}
where $(a)_n=a(a+1)\cdots (a+n-1)$ is called at present the Pochhammer symbol.
Actually, he established this simplest hypergeometric
functions identity for fractional values of $a$ and his main achievement consisted in the treatment
of infinite series.

The major development of the theory of special functions of hypergeometric type took place in the hands of
Leonhard Euler \cite{aar} (I am using this textbook as a key source of historical data).
Among his tremendous list of glorious discoveries one can distinguish the following ones:
from 1729 he sequentially introduced the gamma function $\Gamma(x)$,
\begin{equation}
\Gamma(x)=\int_0^\infty t^{x-1}e^{-t}dt, \qquad \text{Re}(x)>0,
\label{gamma}\end{equation}
the beta function (integral) $B(x,y)$,
\begin{equation}
B(x,y)=\int_0^1t^{x-1}(1-t)^{y-1}dt=\frac{\Gamma(x)\Gamma(y)}
{\Gamma(x+y)}, \qquad \text{Re}(x), \text{Re}(y)>0,
\label{Bfunction}\end{equation}
and the key hypergeometric ${}_2F_1$-function,
\begin{equation}
{}_2F_1(a,b;c;x)
=\sum_{n=0}^\infty \frac{(a)_n(b)_n}{n!(c)_n}x^n,
\qquad |x|<1.
\label{2F1}\end{equation}
The Euler integral representation
\begin{equation}
_2F_1(a,b;c;x)=\frac{\Gamma(c)}{\Gamma(c-b)\Gamma(b)}\int_0^1 t^{b-1}(1-t)^{c-b-1}
(1-xt)^{-a}dt,
\label{2F1int}\end{equation}
where $\text{Re}(c)>\text{Re}(b)>0$ and $x\notin [1,\infty[$, follows directly
from expanding integrand's $(1-xt)^{-a}$ factor into the Taylor series according to formula \eqref{binomial}
and using the exact integration formula \eqref{Bfunction}.

Gauss (1812), Kummer (1836), Riemann (1857), Barnes (1908)
investigated in detail properties of the ${}_2F_1$-function \cite{aar}. In particular,
the hypergeometric equation (considered by Euler already in 1769) satisfied by it:
$$
x(1-x)y''(x)+(c-(a+b+1)x)y'(x)-ab y(x)=0.
$$
The enormous popularity of this equation is explained by the fact that it represents
a very geometric object -- the general differential equation of the second order
with three regular singular points (fixed as $0,1$ and $\infty$). All series expansions
of solutions at these singular points are expressed in terms of the ${}_2F_1$-function
($y(x)= {}_2F_1(a,b;c;x)$ is the solution analytical near the point $x=0$).

As to the special functions of many variables, I will mention only the multiple beta integral
evaluated by Atle Selberg in 1944 \cite{aar}:
\begin{eqnarray}\nonumber &&
\int_0^1 \ldots
\int_0^1 \prod_{1 \leq i < j \leq n} |x_i - x_j|^{2\gamma}
 \prod_{j=1}^n x_j^{\alpha-1} (1 - x_j)^{\beta-1} dx_j
 \nonumber \\
&& \makebox[2em]{}
= \prod_{j=1}^n \frac{\Gamma(\alpha+(j-1)\gamma)
\Gamma(\beta+(j-1)\gamma)
\Gamma(1+j\gamma)}{\Gamma(\alpha+\beta+(N+j-2)\gamma)
\Gamma(1+\gamma)},
\label{selberg} \\ &&
\text{Re}(\alpha ), \text{Re} (\beta) >0, \quad
\text{Re}(\gamma ) > - \min \left(\frac{1}{n},
\frac{\text{Re} (\alpha )}{n-1},
\frac{\text{Re} (\beta)}{n-1}\right).
\nonumber\end{eqnarray}
It was introduced for some number-theoretical needs, but its most important applications have been found
in mathematical physics: random matrices, integrable $n$-body problems, multiple orthogonal polynomials, etc.

{\bf $q$-Hypergeometric functions.} Again, a systematic consideration of the second level of hypergeometric
functions was launched by Euler, who constructed in 1748 the $q$-exponential functions
$$
\sum_{n=0}^\infty \frac{x^n}{(q;q)_n} =\frac{1}{(x;q)_\infty},\quad   |x|<1,\qquad
\sum_{n=0}^\infty \frac{q^{n(n-1)/2}}{(q;q)_n}\; (-x)^n=(x;q)_\infty,\quad |q|<1,
$$
where the product $(x;q)_n=\prod_{k=0}^{n-1} (1-xq^k)$ is known today as the $q$-Pochhammer symbol.
Many mathematicians considered generalizations of these exact summation formulas. In particular,
Rothe, Cauchy, Heine, Gauss have established the following identity
$$
{}_{1}\varphi_0(t;q,x)=\sum_{n=0}^\infty \frac{(t;q)_n}{(q;q)_n}\;x^n
=\frac{(tx;q)_\infty}{(x;q)_\infty}, \quad |x|, |q|<1,
$$
which is referred to as the $q$-binomial theorem.
Heine in 1847 constructed a $q$-analogue of the $_2F_1$-function
$$
{}_{2}\varphi_1(s,t;w;q,x)
=\sum_{n=0}^\infty\frac{(s;q)_n(t;q)_n}{(q;q)_n(w;q)_n}\;x^n,
$$
such that
\begin{equation}
{}_2\varphi_1(q^a,q^b;q^c;q,x)\to {}_2F_1(a,b;c;x) \qquad \text{for} \quad  q\to 1.
\label{limit1}\end{equation}

The theory of hypergeometric functions was developing for more than 300 years in these two instances:
$$
{}_{r+1}F_r\left({u_1,\ldots,u_{r+1} \atop v_1,\ldots, v_r};x\right)
=\sum_{n=0}^\infty\frac{(u_1)_n\cdots (u_{r+1})_n}{n!(v_1)_n\cdots(v_r)_n}\;x^n
$$
and
$$
{}_{r+1}\varphi_r\left({t_1,\ldots, t_{r+1} \atop w_1,\cdots,w_r};q,x\right)
=\sum_{n=0}^\infty \frac{(t_1;q)_n\cdots (t_{r+1};q)_n}{(q;q)_n (w_1;q)_n
\ldots (w_r;q)_n}\;x^n,
$$
together with their multivariable extensions both in the series and integral forms.
In 1980s some people expressed an opinion that no good special functions of hypergeometric
type exist beyond them. Therefore it was a very big surprise when around the turn of Millenium
the following functions have been discovered.

{\bf Elliptic hypergeometric functions \cite{Rosengren,spi:umnrev}.}
What is the top presently known extension of the binomial theorem, Euler beta integral, etc,
and where that has been found?
I was lucky to make one of the major contributions to answering this question during the work at the
Bogolyubov Laboratory of Theoretical Physics, JINR.$^{\! 1}$%
\footnote{$^1$Here it should be mentioned that I am a member of the N. N. Bogolyubov school from the
times of my 1982 post-graduate studies at the Chair of Quantum Statistics and Field Theory of
the Physics Department of the Moscow State University. My PhD thesis supervisors were
V. A. Matveev and K. G. Chetyrkin -- antecedent generation members of this school.}
Namely, in 2000 the elliptic beta integral has been discovered and
its exact evaluation was established as the following theorem \cite{spi:umn}.

{\bf Theorem.}  Let $p, q, t_j\in\mathbb{C},$ $|p|, |q|, |t_j|<1$ and $\prod_{j=1}^6t_j=pq$. Then
\begin{equation}
\frac{(p;p)_\infty(q;q)_\infty}{4\pi \textup{i}}
\int_\mathbb{T}\frac{\prod_{j=1}^6
\Gamma(t_jz^{{\pm 1}};p,q)}{\Gamma(z^{\pm 2};p,q)}\frac{dz}{z}
=\prod_{1\leq j<k\leq6}\Gamma(t_jt_k;p,q),
\label{ellbeta}\end{equation}
where $\mathbb{T}$ is the unit circle and both sides of the equality are composed out of the
elliptic gamma function
$$
\Gamma(z;p,q):=
\prod_{j,k=0}^\infty\frac{1-z^{-1}p^{j+1}q^{k+1}}{1-zp^{j}q^{k}}, \quad |p|, |q|<1,
$$
according to the conventions
$$
\Gamma(t_1,\ldots,t_k;p,q):=\Gamma(t_1;p,q)\cdots\Gamma(t_k;p,q),\quad
\Gamma(tz^{\pm1};p,q):=\Gamma(tz;p,q)\Gamma(tz^{-1};p,q).
$$

The integrand function in \eqref{ellbeta} satisfies a linear $q$-difference equation of the first order
with the coefficient given by a particular elliptic function, which follows from the generating equation
$$
\Gamma(qz;p,q)=\theta(z;p)\Gamma(z;p,q), \quad
\theta(z;p)=(z;p)_\infty(pz^{-1};p)_\infty,
$$
where $\theta(z;p)$ is a Jacobi theta function with a specific normalization
$$
\theta(z;p)=\frac{1}{(p;p)_\infty} \sum_{k\in\mathbb{Z}} (-1)^kp^{k(k-1)/2}z^k.
$$

Relation \eqref{ellbeta} is unique and pretends to be the most important exact evaluation formula
of a univariate integral found so far. This statement is justified by the following facts:

\begin{itemize}

\item
Identity \eqref{ellbeta} is an elliptic analogue of the binomial theorem.

\item
It is the top known univariate extension of the Euler beta integral (which includes also the Gaussian integral).

\item
It defines the measure for two-index biorthogonal functions --
the most general univariate special functions with classical properties \cite{spi:theta}.

\item
It serves as a germ for all elliptic hypergeometric integrals admitting exact evaluation (including
an elliptic analogue of the Selberg integral) and of the whole theory of transcendental
elliptic hypergeometric functions \cite{spi:umnrev}.

\item
It proves the confinement phenomenon in a special sector of states of the simplest $4d$ supersymmetric field theory
(as an equality of superconformal indices of dual theories) \cite{DO}.

\item
It defines the elliptic Fourier transformation \cite{spi:bailey2} with nice
inversion property \cite{SW}.
The key algebraic identity emerging in the corresponding
Bailey lemma is nothing but the star-triangle relation in the operator
form, which coincides with the braiding relation for generators of the
permutation group \cite{DS}.
The functional form of this relation serves as a master identity
for exactly solvable $2d$ spin lattice systems of the Ising type \cite{BS}.

\end{itemize}

\vskip 5mm
\centerline{\bf \scriptsize
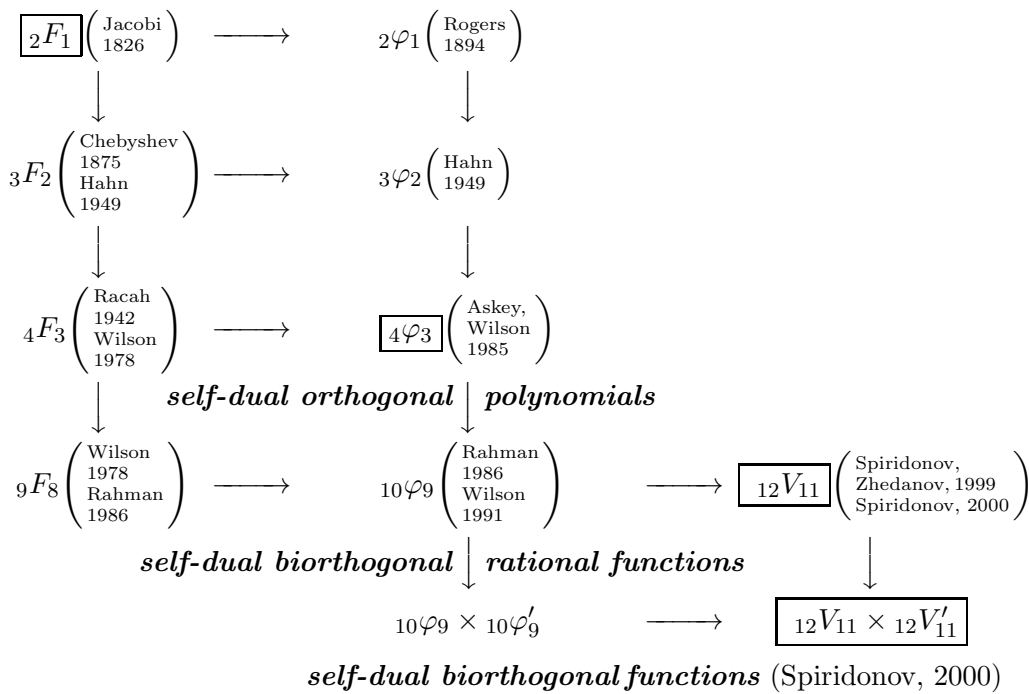
Table. \underline{\bf CLASSICAL (BI)ORTHOGONAL SPECIAL FUNCTIONS}}
$$
\begin{CD}
\text{\hskip 5mm}
\makebox[-1.1em]{}
\fbox{{${}_2F_1$}}
{\tiny\left(\!\!
\begin{array}{l}
\text{Jacobi}\\
 \text{1826}\\
\end{array}\!\!
\right)}
@>>>
{ \makebox[-1.4em]{}\text{
${}_2\varphi_1$}}
{\tiny\left(\!\!
\begin{array}{l}
\text{Rogers} \\
\text{1894}
\end{array}\!\!
\right)}
@.\\
@VVV @VVV@.\\
{ \text{
${}_3F_2$}}
{\tiny\left(\!\!
\begin{array}{l}
\text{Chebyshev}\\
\text{1875}\\
\text{Hahn}\\
\text{1949}
\end{array}\!\!
\right)}
\text{\hskip -1mm}
@>>>
{ \text{
${}_3\varphi_2$}}
{\tiny\left(\!\!
\begin{array}{l}
\text{Hahn}\\
\text{1949}
\end{array}\!\!
\right)}
\text{\hskip 7mm}
@.\\
@VVV @VVV@.\\
{ \text{
${}_4F_3$}}
{\tiny\left(\!\!
\begin{array}{l}
\text{Racah}\\
\text{1942}\\
\text{Wilson}\\
\text{1978}
\end{array}\!\!
\right)}
@>>>
{
\fbox{${}_4\varphi_3$}}
{\tiny\left(\!\!
\begin{array}{l}
\text{Askey,}\\
\text{Wilson}\\
\text{1985}
\end{array}\!\!
\right)}
@.\\
@VVV @V{{\text{{\normalsize \bf \em  self-dual orthogonal}}\atop }\atop}
V
{{\text{{\normalsize  \bf\em polynomials}}\atop }\atop}
V@.\\
{ \text{
${}_9F_8$}}
{\tiny\left(\!\!
\begin{array}{l}
\text{Wilson}\\
\text{1978}\\
\text{Rahman}\\
\text{1986}
\end{array}\!\!
\right)}
@>>>
{ \text{
${}_{10}\varphi_9$}}
{\tiny\left(\!\!
\begin{array}{l}
\text{Rahman}\\
\text{1986}\\
\text{Wilson}\\
\text{1991}
\end{array}\!\!
\right)}
@>>>
\fbox{{ ${}_{12}V_{11}$}}
{\tiny\left(\!\!
\begin{array}{l}
\text{Spiridonov,}\\
\text{Zhedanov},
\text{1999}\\
\text{Spiridonov, 2000}
\end{array}\!\!
\right)}
@.\\
@. \makebox[-11em]{}  \text{\normalsize \bf \em  self-dual
biorthogonal} $\;$ @VVV
@. \makebox[-20em]{}  \text{\normalsize \bf \em \makebox[-4em]{} rational functions $\quad$ }
$\qquad$
 @   VVV \\
@.
 {}_{10}\varphi_{9}\times {}_{10}\varphi_{9}'
@>>>
\makebox[-1em]{} \fbox{{ ${}_{12}V_{11}\times {}_{12}V_{11}'$}} @.  \\
@. \text{\normalsize {\bf \em  self-dual
biorthogonal
}} @. \makebox[-5em]{} \text{\normalsize {\bf \em
functions} (Spiridonov, 2000)} @.  \\
\end{CD}
$$

\vskip 5mm

{\bf Classical special functions.}
Let us unfold some of the arguments given above. First we give explicit definition of the very-well-poised
elliptic hypergeometric series \cite{spi:umnrev}:
\begin{equation}
{}_{r+1}V_{r}(t_0;t_1,\ldots,t_{r-4};q,p)
= \sum_{n=0}^\infty \frac{\theta(t_0q^{2n};p)}{\theta(t_0;p)}\prod_{m=0}^{r-4}
\frac{\theta(t_m)_n}{\theta(qt_0t_m^{-1})_n}q^n,
\label{eseries} \end{equation}
where $\theta(z)_n=\prod_{k=0}^{n-1}\theta(zq^k;p)$ is the elliptic Pochhammer symbol.
The word combination ``elliptic series'' in this context means that the parameters in \eqref{eseries}
satisfy the balancing condition $\prod_{k=1}^{r-4}t_k=t_0^{(r-5)/2}q^{(r-7)/2},$
which guarantees that each term of this series is an elliptic function (a meromorphic
double periodic function) of all its parameters.
However, the infinite sum of elliptic functions \eqref{eseries} in general does not converge.
Therefore one has to terminate it by imposing the constraint $t_j=q^{-N},\, N=0,1,\ldots$,
for some fixed $j$.
Similar to the limit \eqref{limit1}, for fixed parameters $t_m$
\begin{equation}
 \lim_{p\to 0} {}_{r+1}V_r = \text{ very-well poised, balanced}
\  {}_{r-1}\varphi_{r-2}\text{-series}.
\label{limit2}\end{equation}

The Frenkel-Turaev sum \cite{FT} provides a closed form expression for the terminating $_{10}V_9$-series
and it is a special limiting case of the elliptic beta integral \eqref{ellbeta}.
The terminating $_{12}V_{11}$-series emerges in solutions of the IRF type Yang-Baxter
equation \cite{FT} and of the Lax pair equations for a $2d$ discrete-time chain generalizaing
the Toda lattice \cite{SZ}. The set of classical (bi)orthogonal functions is described
in the enclosed Table -- an extension of the Askey scheme for orthogonal polynomials.
Its top left corner belongs to the Jacobi polynomials
defined by the terminating $_2F_1$-series \eqref{2F1}, which are orthogonal with respect to the
measure \eqref{Bfunction}. And that is the end of differential equations in this
context. The most general classical orthogonal polynomials were found
by Askey and Wilson \cite{AW} and they are defined by a terminating $_4\varphi_3$-series.
Self-duality means that these polynomials satisfy finite-difference equations of the
second order both in the degree of polynomials and in their argument with the same coefficients
(i.e. there is a permutational symmetry between the corresponding variables).

Elliptic analogues of the plain and $q$-hypergeometric classical special functions naturally
emerge only at the top $q$-hypergeometric level. Namely, two particular $_{12}V_{11}$ terminating
series form a pair of biorthogonal rational functions representing an
elliptic extension of the $q$-Racah polynomials, as described in \cite{SZ}. Proper
generalization of the Askey-Wilson polynomials was established in \cite{spi:theta}.
Moreover, a principally new phenomenon of two-index biorthogonality of (non-rational!)
functions emerges at this level, as indicated in the right bottom
corner of the Table. The latter top classical functions were also discovered
at BLTP JINR \cite{spi:theta}. The presented scheme is far from complete since not all
possible limiting transitions and potential generalizations are depicted in it.

{\bf Superconformal index.} Four-dimensional minimal superconformal quantum field theories are
based on the full symmetry group $G_{st}\times G\times F$, where $G_{st}=SU(2,2|1)$ is the
flat space-time symmetry group, $G$ is the local gauge invariance group, and $F$ is the flavor
group of global internal symmetries. The $SU(2,2|1)$ supergroup  is generated by
$J_i, \overline{J}_i$ ($SL(2,\mathbb{C})$ group generators, or Lorentz
rotations), $P_\mu, Q_{\alpha},\overline{Q}_{\dot\alpha}$ (supertranslations),
$K_\mu, S_{\alpha},\overline{S}_{\dot\alpha}$ (special superconformal transformations),
$H$ (dilations) and $R$ ($U(1)_R$-rotations).
Pick up a distinguished pair of supercharges, e.g.,
$Q\propto \overline{Q}_{1 }$, $Q^{\dag}\propto {\overline S}_{1}$
and the maximal Cartan subalgebra generators commuting with them, $H-R/2,\, J_3$, and
$F_k$ (maximal torus generators of $F$). Then one has
\begin{equation}
Q^2=(Q^{\dag})^2=0,\quad \{Q,Q^{\dag}\}= 2{\mathcal H},\qquad \mathcal{H}=H-2\overline{J}_3-3R/2,
\label{susy}\end{equation}
and the superconformal index is defined as the trace of the following operator \cite{Kinney,Romelsberger}
\begin{equation}
I(p,q,y_k) = \text{Tr} \Big( (-1)^{\rm F}
p^{\mathcal{R}/2+J_3}q^{\mathcal{R}/2-J_3} e^{-\beta {\mathcal H}} \prod y_k^{F_k}\Big),
\label{Ind}\end{equation}
where $\mathcal{R}= H-R/2$, $(-1)^{\rm F}$ is the $\mathbb{Z}_2$-grading operator, and
$p,q,y_k$ are fugacities (group parameters). The trace may get non-zero contributions only from
the BPS states $Q|\psi\rangle=Q^{\dag}|\psi\rangle=\mathcal{H}|\psi\rangle=0$,
so that the $\beta$-dependence is cancelled out.
A heuristic computation yields the matrix integral
\begin{equation}
I(y;p,q) \ = \ \int_{G} d \mu(z)\,  \exp \bigg ( \sum_{n=1}^{\infty}
\frac 1n ind\big(p^n ,q^n, z^n , y^ n\big ) \bigg),
\label{IndExplicit}\end{equation}
where  $\mu(z)$ is the Haar measure for group $G$ and
\begin{eqnarray}\nonumber &&
ind(p,q,z,y) =  \frac{2pq - p - q}{(1-p)(1-q)} \chi_{adj_G}(z)\cr
\\ && \makebox[2em]{}
+ \sum_j \frac{(pq)^{r_j}\chi_{R_F,j}(y)\chi_{R_G,j}(z) - (pq)^{1-r_j}
\chi_{{\bar R}_F,j}(y)\chi_{{\bar R}_G,j}(z)}{(1-p)(1-q)}.
\label{1index}\end{eqnarray}
Here $\chi_{R_F,j}(y)$, $\chi_{{\bar R}_F,j}(y)$, and $\chi_{R_G,j}(z)$ are the group characters
of respective field representations and $2r_j\in\mathbb{Q}$ are their $R$-charges.

For the unitary group $SU(N)$,
$z=(z_1,\ldots,z_N), \ \prod_{a=1}^Nz_a=1$,
\begin{eqnarray*}
\int_{SU(N)} d\mu(z) \ = \   \frac{1}{N!} \int_{\mathbb{T}^{N-1}}
\Delta(z) \Delta(z^{-1}) \prod_{a=1}^{N-1} \frac{dz_a}{2 \pi \textup{i} z_a},
\quad
\Delta(z) = \prod_{1 \leq a < b \leq N} (z_a-z_b).
\end{eqnarray*}

{\bf Seiberg duality.} Consider electromagnetic duality of the following
$4d$ $\mathcal{N}=1$ supersymmetric field theories conjectured by Seiberg in 1994 \cite{Seiberg}.

{\em Electric theory} (weak coupling regime): $G=SU(2),$ $F=SU(6)$, the field/representation content
\begin{eqnarray*} &&
1)\ \text{vector superfield}:  (adj, 1), \qquad \chi_{SU(2),adj}(z)=z^2+z^{-2}+1,
\\ &&
2)\ \text{chiral superfield}: (f, f),\qquad \chi_{SU(2),f}(z)=z+z^{-1},\qquad r_f=1/6,
\\ && \makebox[0em]{} 
\chi_{SU(6),f}(y)=\sum_{k=1}^6y_k,
\quad \chi_{SU(6),\bar f}(y)=\sum_{k=1}^6y_k^{-1},
\quad \prod_{k=1}^6y_k=1.
\end{eqnarray*}

{\em Magnetic theory} (strong coupling): $G=1,$ $F=SU(6)$ with the single field/representation
$T_A:\; \Phi_{ij}=-\Phi_{ji},$
$$
\chi_{SU(6),T_A}(y)=\sum_{1\leq i<j\leq 6}y_iy_j,
\qquad r_{T_A}=1/3.
$$

A relation of these theories to  the elliptic beta integral was discovered by Dolan and Osborn
in 2008 \cite{DO}. Namely, after explicit computation of superconformal indices in these theories,
$I_E$ and $I_M$, and equating them according to the Seiberg duality conjecture,
there emerges precisely the elliptic beta integral evaluation formula \eqref{ellbeta} in the form
$$
I_E\; (the\; l.h.s.) = I_M\; (the\; r.h.s.)
$$
with the identification of parameters $t_k=(pq)^{1/6}y_k, \; k=1,\ldots, 6.$

In general, the explicit computability of elliptic hypergeometric integrals serves
as the confinement criterion in $4d$ $\mathcal{N}=1$ supersymmetric gauge theories.
The process of integrals' evaluation has the physical meaning of a transition from
the weak to strong coupling regimes in quantum field theories. Among
conceptual interpretations of the exact mathematical formulas of the type $``A=B"$
this example is perhaps the brightest one.

General Seiberg duality \cite{Seiberg} deals with much more complicated
field theories described in the tables below, where adj, $f$ and $\bar f$ mean adjoint,
fundamental and antifundamental representations, and the last two columns contain
corresponding $U(1)$-groups charge values.

{\em Electric theory} ($G=SU(N),\, F=SU(M )_l\times SU(M )_r\times U(1)_B$, $\tilde N =M -N$):
\begin{center}\begin{tabular}{|c|c|c|c|c|}
\hline
 $SU(N )$ & $SU(M )_l$ & $SU(M )_r$ & $U(1)_B$ & $U(1)_R$ \\
\hline
 $f$ & $f$ & 1 & 1 & $\tilde{N }/M $ \\
 $\overline{f}$ & 1 & $\overline{f}$ & -1 & $\tilde{N }/M $ \\
 ${\rm adj}$  & $1$   &  $1$ &  $0$   &  $1$ \\
\hline
\end{tabular}
\end{center}

{\em Magnetic theory} ($G=SU(\tilde N), \, F$ is the same):
\begin{center}
\begin{tabular}{|c|c|c|c|c|}
\hline
 $SU(\tilde N )$ & $SU(M )_l$ & $SU(M )_r$ & $U(1)_B$ & $U(1)_R$ \\
\hline
 $f$ & $\overline{f}$ & 1 & $N /\tilde N $ & $N /M $ \\
 $\overline{f}$ & 1 & $f$ & $-N /\tilde N $ & $N /M $
\\
 1 & $f$ & $\overline{f}$ & 0 & $2\tilde N /M $
\\
 ${\rm adj}$  & $1$   &  $1$ &  $0$   &  $1$ \\
\hline
\end{tabular}
\end{center}
Seiberg conjectured that these two non-abelian gauge field theories are equivalent at the infrared fixed
points for $3N /2 < M  < 3 N $ (the conformal window). This conjecture was proved by Dolan and
Osborn \cite{DO} in the sector of BPS states using the mathematical theorems on symmetry
properties of particular elliptic hypergeometric integrals. So, in proper parametrization,
the electric index takes the form
\begin{eqnarray}\nonumber
&& I_E = \kappa_{N } \int_{\mathbb{T}^{N -1}}
\frac{\prod_{i=1}^{M } \prod_{j=1}^{N } \Gamma(s_i z_j,t^{-1}_i
z^{-1}_j;p,q)}{\prod_{1 \leq i < j \leq N } \Gamma(z_i
z^{-1}_j,z_i^{-1} z_j;p,q)} \prod_{j=1}^{N -1} \frac{d z_j}{2 \pi \textup{i}
z_j},
\end{eqnarray}
where $\prod_{j=1}^{N } z_j =1,\, \kappa_N=(p;p)_\infty^{N-1} (q;q)_\infty^{N-1}/N!.$
The magnetic index takes the form
\begin{eqnarray}\nonumber &&
I_M = \kappa_{\tilde N }\prod_{i,j=1}^{M } \Gamma(s_i t^{-1}_j;p,q)
  \int_{\mathbb{T}^{\widetilde N -1}}
\frac{\prod_{i=1}^{M } \prod_{j=1}^{\widetilde N }
\Gamma(S^{\frac{1}{\widetilde N }} s_i^{-1}
x_j,T^{-\frac{1}{\widetilde N }} t_i x_j^{-1};p,q)}{\prod_{1 \leq i <
j \leq \widetilde N } \Gamma(x_i
x_j^{-1},x_i^{-1}x_j;p,q)}\prod_{j=1}^{\widetilde N -1}  \frac{dx_j}{2 \pi \textup{i} x_j},
\nonumber\end{eqnarray}
where $ \prod_{j=1}^{\tilde N } x_j =1$, $S=\prod_{i=1}^{M } s_i,\,  T = \prod_{i=1}^{M } t_i, \,
ST^{-1} = (pq)^{M -N }.$

The equality $I_E=I_M$ in some particular cases has been established or conjectured
in my papers \cite{spi:umn,spi:theta} and it was proved in full generality by Rains in \cite{rai:trans}.

The elliptic analogue of the Selberg integral has been introduced in \cite{vDS1}
and its relation to the Seiberg type dualities was described in \cite{SV}.
For $|p|, |q|, |t|, |t_m|<1$ and $t^{2n-2}\prod_{m=1}^6t_m=pq$,
\begin{eqnarray}\nonumber
&& \makebox[-1em]{}
\frac{(p;p)_\infty^n (q;q)_\infty^n }{ (4\pi\textup{i})^n n!}
\int_{\mathbb{T}^n} \prod_{1\leq j<k\leq n}
\frac{\Gamma(tz_j^{\pm 1} z_k^{\pm 1};p,q)}{\Gamma(z_j^{\pm 1} z_k^{\pm 1};p,q)}
\prod_{j=1}^n\frac{\prod_{m=1}^6\Gamma(t_mz_j^{\pm 1};p,q)}{\Gamma(z_j^{\pm2};p,q)}
\frac{dz_j}{z_j}
\\  && \makebox[2em]{}
=\prod_{j=1}^n\left(\frac{\Gamma(t^j;p,q)}{\Gamma(t;p,q)}
\prod_{1\leq m<s\leq 6}\Gamma(t^{j-1}t_mt_s;p,q)\right).
\label{eSelberg}\end{eqnarray}

A systematic analysis \cite{SV} has shown that the physics of Seiberg like dualities
yields many new complicated mathematical conjectures (special function identities), while the
mathematics of elliptic hypergeometric integrals
produces many new electromagnetic dualities. In a sense, on this ground physics and
mathematics work again hand-to-hand as one science. As another important physical outcome,
I would like to mention that
superconformal indices of quiver gauge theories describe partition functions of $2d$ spin systems of
Ising type, where  Seiberg duality serves as the integrability condition \cite{stat}.

Nowadays, computations of supersymmetric partition functions became an industry in
producing special function identities with many contributors \cite{RR2016}. Let me mention
some other results: the work \cite{GPRR} deals with applications to $2d$ topological field theories,
the work \cite{DG} deals with a physical interpretation of $W(E_7)$ symmetry of the elliptic extension
of Euler-Gauss hypergeometric function \cite{spi:umnrev}, in the paper \cite{GK} an intriguing hypothesis
was put forward on the relation to Nekrasov instanton sums, \cite{ACM} contains an exact computation of
the $4d$ supersymmetric partition functions by the localization technique, in an interesting paper
\cite{Yagi} a deeper picture of relations to $2d$ solvable statistical mechanics systems is given,
some more recent developments of the subject are reflected in the papers \cite{CKW,SS}.

{\em Right after giving this talk I have written my last message to Dick as a part of the Liber Amicorum
collection prepared by his numerous friends and shortly handed to him, which was a very timely step.}

\end{document}